# High Accuracy Classification of Parkinson's Disease through Shape Analysis and Surface Fitting in $^{123}$I-Ioflupane SPECT Imaging

R. Prashanth, *Student Member, IEEE*, S. Dutta Roy, *Member, IEEE*, Pravat K. Mandal, and S. Ghosh

*Abstract*—**Early and accurate identification of parkinsonian syndromes (PS) involving presynaptic degeneration from non-degenerative variants such as Scans Without Evidence of Dopaminergic Deficit (SWEDD) and tremor disorders, is important for effective patient management as the course, therapy and prognosis differ substantially between the two groups. In this study, we use Single Photon Emission Computed Tomography (SPECT) images from healthy normal, early PD and SWEDD subjects, as obtained from the Parkinson's Progression Markers Initiative (PPMI) database, and process them to compute shape- and surface fitting-based features for the three groups. We use these features to develop and compare various classification models that can discriminate between scans showing dopaminergic deficit, as in PD, from scans without the deficit, as in healthy normal or SWEDD. Along with it, we also compare these features with Striatal Binding Ratio (SBR)-based features, which are well-established and clinically used, by computing a feature importance score using Random forests technique. We observe that the Support Vector Machine (SVM) classifier gave the best performance with an accuracy of 97.29%. These features also showed higher importance than the SBR-based features. We infer from the study that shape analysis and surface fitting are useful and promising methods for extracting discriminatory features that can be used to develop diagnostic models that might have the potential to help clinicians in the diagnostic process.**

## I. INTRODUCTION

PARKINSONIAN SYNDROMES (PS) is a group of movement disorders that is clinically characterized by symptoms of resting tremor, rigidity and bradykinesia [1]. The clinical diagnosis of PD based on clinical signs and a good response to levodopa, can be straightforward. However, in the early stages of the disease, when the symptoms are mild, atypical or ambiguous with unconvincing responses to levodopa, the diagnosis can be difficult and inconclusive [1, 2]. The Parkinson's Progression Markers Initiative (PPMI) [3] which is a landmark large-scale study to identify PD progression biomarkers points out that early diagnosis of *de novo* PD subjects, like those being recruited for the study, is difficult because the characteristic signs and symptoms of the disease have not yet fully emerged (Study Protocol of the PPMI; http://www.ppmi-info.org/study-design/research-documents-and-sops/). Few tremor disorders, such as Essential Tremor (ET) that do not depict any dopaminergic deficit, but share several clinical features as in PS, can also lead to difficulties in the diagnostic process [4].

Early and accurate diagnosis of PS involving presynaptic degeneration is of prime importance for effective disease management and for allowing neuroprotective strategies to be administered earlier when they become available [2]. Accurate identification is crucial for effective patient management because the disease course, prognosis and therapy differ substantially from the non-degenerative variants or other tremor disorders [4].

SPECT imaging using $^{123}$I-Ioflupane (DaTSCAN or [123I]FP-CIT) is presently among the most sensitive imaging techniques, even in the early stages of the disease [1, 2, 5]. Dopaminergic imaging discriminates patients with neurodegenerative PS from healthy normal, non-degenerative PS and tremor disorders such as ET by identifying presynaptic dopaminergic deficits in the caudate and putamen with high sensitivity and specificity [2]. Based on the pattern of uptake of the radiotracer, SPECT images can be normal (that shows no dopaminergic deficit) or abnormal. Normal scans are characterized by intense and symmetric DAT binding in the caudate nucleus and putamen on both hemispheres that appear as two 'comma' shaped regions (Figs. 2(a) and 2(e)). Any asymmetry or distortion of this shape implies an abnormal finding (Fig. 2(i)) [1, 4]. A number of studies on early PD have observed that about 10–15% of subjects recruited in their studies by movement disorder experts with the diagnosis of PD, showed Scans Without Evidence of Dopaminergic Deficit or normal dopaminergic activity, which led to the coining of the term SWEDD [6-8]. Subsequent follow-up showed that they neither deteriorate nor respond to levodopa, and that their SPECT scans remain normal [8, 9]. It was inferred that they

R. Prashanth* was with the Department of Electrical Engineering, Indian Institute of Technology Delhi, New Delhi 110016, India (*corresponding author, Email: prashanth.r.iitd@gmail.com, Ph: +91-9891279885)

S. Dutta Roy is with the Department of Electrical Engineering, Indian Institute of Technology Delhi, New Delhi 110016, India; Email: sumantra@ee.iitd.ac.in).

Pravat K. Mandal is with the Neuroimaging and Neurospectroscopy Lab, National Brain Research Centre, Manesar 122050, India (pravat@nbrc.ac.in).

S. Ghosh was with the Martinos Center for Biomedical Imaging, Massachusetts General Hospital and Harvard Medical School, Boston, Massachusetts, USA (shantanu@nmr.mgh.harvard.edu).

were highly unlikely of having PD. The results from these studies clearly point out that dopaminergic imaging is highly useful and that an abnormal imaging, at least in cases of diagnostic uncertainty, is strongly supportive of a diagnosis of neurodegenerative PS.

In clinical practice, SPECT images are usually evaluated visually or through region-of-interest (ROI) analysis [10]. Researchers have also carried out voxel-based analysis where voxel clusters that show significant decrease in the uptake are identified [10-14]. Visual analysis, however, relies on the judgement of the observer that heavily depends on his expertise and knowledge [10]. ROI techniques involve outlining or positioning the ROI over the striatum (target region) and the occipital cortex (reference region), and a quantitative measure termed the background subtracted striatal uptake ratio is computed [10]. Despite odds, the quantitative method is the most acceptable one, since, according to Phase III trial, it provides an excellent intra- and inter-observer agreement. Visual assessment may lead to pitfalls. PPMI provides quantified striatal values, called the Striatal Binding Ratio (SBR) values, via their database and they are computed by nuclear medicine experts of the PPMI (SPECT Manual of the PPMI; http://www.ppmi-info.org/study-design/research-documents-and-sops/). These quantitative measures may be more helpful in cases when there is ambiguity in visual assessment [15]. On the other hand, voxel-based techniques are widely used for scientific purposes but are observed to be not practical for use in routine clinical practise [16].

An alternate approach is to carry out shape and intensity distribution (surface profile) analysis, and use pattern recognition techniques for differentiation. This method has the following advantages: 1) It does not require positioning of ROI, 2) It can be automated or semi-automated thus, avoiding or reducing inter-operator and intra-operator variability, 3) Shape metric is more strongly associated with the visual appearance of the striatal uptake than the striatal uptake ratio measurement [15]. To the best of our knowledge, there has been only one study [15] which carry out shape analysis to assess patients with PS. They observed that it is a viable alternative to conventional techniques for analysing SPECT images. They segmented regions corresponding to higher uptake areas of striatum and quantify the shape by fitting an ellipse to the region, followed by computing the aspect ratio of the fitted ellipse (they called it as north/south, east/west ratio or NSEW ratio in the paper). Although the shape could be quantified with many different parameters, they had limited to using only this one parameter. The other limitation of the study is that they had limited sample size of 52 subjects (27 with neurodegenerative PS, and 25 had normal scans that include healthy normal and subjects with non-degenerative PS or movement disorders such as ET).

Surface fitting is a useful approach that is widely used in biomedical applications [17-19]. Surface fitting using implicit polynomial functions of degree greater than two can efficiently represent surfaces that are more complicated than those represented by quadric surfaces (e.g., ellipsoid, paraboloid, etc.) [20]. As the intensity distribution in the uptake regions also vary during diseased condition, this approach can be useful in extracting discriminatory features from the distribution pattern.

Realizing the potential of these techniques, we had previously carried out shape analysis [21] and surface fitting [22] to extract discriminatory features using a small dataset. In this work, we combine and extend them by using a larger dataset, compute more relevant shape-based features (such as features based on asymmetry of uptake), develop classification models, and compare these features with SBR-based features that are clinically used. Overall, the study can be summarised as follows. We use SPECT scan data of early PD, SWEDD and healthy normal subjects and process these images to segment the regions of high activity. This is followed by two kinds of analysis to extract the features 1) Shape analysis of these regions by computing various shape-based features 2) Fitting of a cubic surface based on the intensities in these segmented regions. We use these features to develop classification models to classify degenerative PS (early PD group in our study) from healthy normal/non-degenerative condition (Normal/SWEDD group) using machine learning techniques. Along with this, we also compare these features with the SBR-based features in a feature importance estimation framework using the Random Forests technique.

## II. MATERIALS AND METHODS

### A. Database and cohort details

Data used in the preparation of this article were obtained from the PPMI database (www.ppmi-info.org/data). For up-to-date information on the study, please visit www.ppmi-info.org. PPMI is a landmark, large-scale, international and multi-centre study to identify PD progression biomarkers [3].

We use SPECT imaging data corresponding to the subject's screening visit, from the database. The images were downloaded on 25[th] June 2014. The numbers of subjects in the study are 208 healthy normal, 427 PD and 80 SWEDD. All the PD patients are in the early stage (Hoehn and Yahr (HY) stage 1 or 2 with mean $\pm$ SD as $1.50 \pm 0.50$) and all the SWEDD subjects (these are the newly diagnosed PD patients based on clinical symptoms, but show normal dopaminergic imaging) show early stage (mean $\pm$ SD HY stage as $1.46 \pm 0.53$) PD symptoms.

### B. Image analysis and feature extraction

#### 1) Preprocessing by PPMI

All SPECT scan data acquired at the PPMI sites undergo a pre-processing procedure before they are publically shared via the database. This pre-processing ensures that all scans were in the same anatomical alignment (spatially normalized). The process include reconstruction from raw projection data, attenuation correction, followed by applying a standard Gaussian 3D 6.0 mm filter, and then normalizing these images to standard Montreal Neurologic Institute (MNI) space [23]. We use these pre-processed scans for analysis and the analysis pipeline is as shown in Fig. 1.

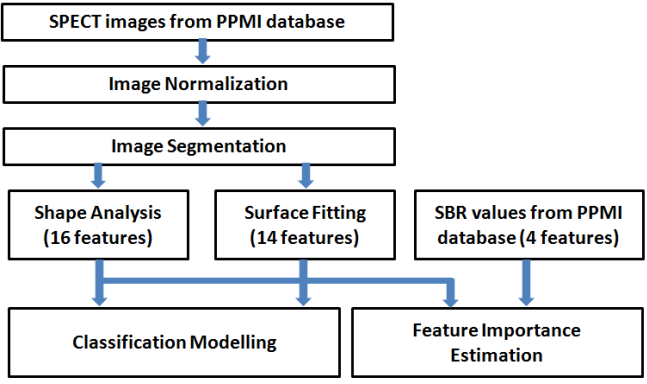

Fig. 1. Processing pipeline

side region is flipped from right to left.

### 5) Surface fitting based on intensity distribution

A polynomial surface is fit based on the intensity distribution in the segmented region. We choose polynomial model of order 3 (cubic model) for surface fitting as higher orders can lead to over-fitting. The cubic model is given by

$$f(x,y) = p_{00} + p_{10}x + p_{01}y + p_{20}x^2 + p_{11}xy + p_{02}y^2 \\ + p_{30}x^3 + p_{12}xy^2 + p_{21}x^2y + p_{03}y^3 \quad (1)$$

where $p_{ij}; i,j \in \{0,1,2,3\}$ are the model coefficients which are estimated using linear least-squares method where it minimizes the summed square of residuals (or errors). Prior to the fitting process, the coordinates $\{[x_i, y_i]; i = 1,2, ..., n\}$ of the pixels in the segmented region is normalized by centering them to zero mean and scaling to unit standard deviation. This transformation won't change the fit theoretically, but it will make the results better conditioned on a computer with finite precision. This removes any scaling problems that may arise. The pixel size is approximately 2 x 2 mm.

The goodness-of- fit of the final model is evaluated using Sum of squares due to Error (SE), $R^2$, Adjusted $R^2$ ($R^2_{adj}$) and Root Mean Squared Error (RMSE) measures.

### 6) Feature set

The 16 features through shape analysis and 14 features (10 model coefficients and 4 goodness-of-fit evaluation measures) via surface fitting form the feature set. Along with these, we compute 4 Striatal Binding Ratio (SBR)-based features using the SBR values of the four striatal regions (left and right caudate, and left and right putamen) which are available from the PPMI database. The SBR based features are caudate SBR, putamen SBR, caudate SBR asymmetry index (AI), putamen AI. We use these SBR-based features for comparison with the shape- and surface fitting-based features. More details are provided in the supplementary document (Table S.II).

From our previous studies [21, 22, 24], we observe that shape-, surface fitting- and SBR-based features show good variations between early PD (who show dopaminergic deficit) and healthy normal or SWEDD (who do not show dopaminergic deficit). While comparing the healthy normal and SWEDD groups, the features do not show substantial variation [21, 22]. These observations are consistent with the dopaminergic imaging perspective for early PD, healthy normal and SWEDD groups [2].

### C. Statistical Analysis of Features

The features are tested for statistical significance using Wilcoxon rank sum test. Statistically significant (*p*-value<0.05) features are used for classification modeling to distinguish scans with deficit from the scans without deficit. Healthy normal and SWEDD groups show similar characteristics on dopaminergic imaging (*p*-value>0.05, Table I).We consider them as a single entity as Normal/SWEDD group for further analysis.

### D. Classification of degenerative PS (early PD) from non-degenerative types (Normal/SWEDD)

We carry out binary classification (early PD vs. Normal/SWEDD) using Support Vector Machine (SVM) [25],

### 2) Selection of slices for processing

A SPECT scan consists of 91 transaxial slices (from bottom to top of the head). The radioligand [123I]FP-CIT specifically binds to the striatal DAT, and with dopaminergic neurodegeneration, DAT density decreases and therefore striatal [123I]FP-CIT uptake is reduced. Therefore, striatum is the region to look for to observe deterioration in DAT imaging. We observe that the slice with the highest striatal uptake is near to the 42nd slice. We selected the slices around it, from 35th to 48th, for each scan for further processing as they depict the dopaminergic activity of the subject. A further discussion on the basis of slice selection in given in Sec. III.D.

### 3) Intensity normalization and Segmentation

Each slice is normalized to the range [0, 1], and then a mean image is generated for a subject. This image is again normalized to [0, 1].

In the next step, image segmentation is carried out to extract the high uptake regions, which correspond to the dopaminergic activity. This is done by converting it to a binary image based on a threshold. The threshold was carefully chosen for each image based on empirical experiments (by observing closely at the accurateness of the segmented regions) (Fig. 2). A further discussion on the threshold selection is given in Sec. III.D. Following this, we perform two kinds of analysis which are 1) Quantification of these regions through shape analysis, and 2) Surface fitting. They are described as below.

### 4) Shape analysis using shape-based features

During the course of most degenerative PS, dopamine transporters are first lost in the putamen and then in the caudate, giving a deficit that proceeds from the posterior to anterior striatum (Fig. 1(i)) [15]. Visually, it can be observed that the shape of the high activity region changes from 'comma' to 'dot' shaped. Asymmetry of the uptake regions in the two hemispheres is also a usual observation [1].

In this paper, we segment these uptake regions and quantify them using various shape-based features. They are area, major axis length, minor axis length, aspect ratio, eccentricity, equivalent diameter, orientation, roundness, area asymmetry index (AI), major axis length AI, minor axis length AI, aspect ratio AI, eccentricity AI, equivalent diameter AI, orientation AI, roundness AI. A description of these features is given in supplementary document (Table S.I). To compute these features, the left side region is kept as reference and the right

Boosted Trees [26], Random Forests [27] and Naïve Bayes [28] techniques. We used LIBSVM library [29] for classification using SVM, statistics toolbox in MATLAB for classification using Naïve Bayes, Boosted Trees and Random Forests. The classifiers are evaluated based on 10-fold cross validation that is repeated 100 times.

### 1) Feature importance estimation

Along with classification, the Random forests [27] technique can also carry out feature importance estimation. In this technique, while choosing $n$ out of $n$ observations with replacement, it omits on average 37% of observations for each decision tree. These are 'out-of-bag' observations. Out-of-bag estimates of feature importance are computed by randomly permuting out-of-bag data across one feature at a time, and estimating the increase in the out-of-bag error due to this permutation. The larger the increase in error, higher the importance of the feature.

## III. RESULTS AND DISCUSSION

Fig. 2 illustrates average SPECT image generated, image segmentation and surface fitting for a healthy normal, SWEDD and early PD subject. It is observed that the uptake regions in both normal and SWEDD groups are 'comma' shaped, whereas in PD (early stage) the region deteriorates to become 'circular' or 'dot' shaped. These shapes are consistent with the clinical perspective of the respective conditions. SWEDD subjects show normal dopaminergic imaging, whereas in PD, the change in shape is due to the deterioration or loss of striatal DATs.

The fitted 3D surfaces for both the left and right side regions are similar for normal (Figs. 4(c & d)) and SWEDD (Figs. 4(g & h)) groups, indicating a symmetric behavior of intensity distributions in these regions. On the other hand, the surfaces for PD (Figs. 4(k & l)) show variation between the left and the right side, indicating an asymmetry in the intensity distributions. Another interesting observation is that the cubic surfaces for PD show more positive curvature, whereas for normal and SWEDD, it shows a saddle-shaped or surface with a negative curvature. This is because during the course of PD, DATs are first lost in the putamen (lower portion of the segmented regions or striatum), then in the caudate (upper portion of the segmented regions), or in other words, it proceeds from posterior to anterior striatum [15]. This loss of DATs in PD in the putamen (or posterior striatum) leads to the loss of negative curvature. Table I lists the values of shape-based (1 to 16), surface fitting-based (17 to 30) and SBR-based (31 to 34) features used in the study.

### A. Feature values, statistical analysis and justification for considering normal and SWEDD as a single group

Clinically, it is established that SWEDD subjects show dopaminergic imaging characteristics similar to that of healthy normal. Table I which shows the values along with the results of the statistical testing of the features, are consistent with this perspective. Box plots of the computed features also indicate the same (included in the supplementary file).

TABLE I
SHAPE-, SURFACE FITTING- AND SBR-BASED FEATURE VALUES (MEAN ± SD), AND STATISTICAL TESTING FOR COMPARING HEALTHY NORMAL VS. SWEDD AND EARLY PD VS. NORMAL/SWEDD

| SNo. | Features | Normal | SWEDD | Early PD | $p_1$ | $p_2$ |
|---|---|---|---|---|---|---|
| 1. | Area | 122.2 ± 18.9 | 123.9 ± 16.7 | 71.8 ± 17.6 | 0.43 | ≈ 0 |
| 2. | Major Axis Length | 16.58 ± 1.47 | 16.65 ± 1.4 | 11.06 ± 1.59 | 0.64 | ≈ 0 |
| 3. | Minor axis length | 9.67 ± 0.78 | 9.75 ± 0.66 | 8.25 ± 1.1 | 0.52 | ≈ 0 |
| 4. | Aspect Ratio | 1.72 ± 0.13 | 1.71 ± 0.12 | 1.35 ± 0.15 | 0.54 | ≈ 0 |
| 5. | Eccentricity | 0.81 ± 0.03 | 0.81 ± 0.03 | 0.63 ± 0.1 | 0.55 | ≈ 0 |
| 6. | Equivalent diameter | 12.43 ± 0.98 | 12.52 ± 0.86 | 9.41 ± 1.21 | 0.45 | ≈ 0 |
| 7. | Orientation | 49.46 ± 7.89 | 50.75 ± 7 | 29.66 ± 17.34 | 0.26 | ≈ 0 |
| 8. | Roundness | 0.88 ± 0.04 | 0.89 ± 0.04 | 1.03 ± 0.06 | 0.8 | ≈ 0 |
| 9. | Area AI* | 0.07 ± 0.06 | 0.09 ± 0.1 | 0.43 ± 0.28 | 0.75 | ≈ 0 |
| 10. | Major Axis Length AI | 0.06 ± 0.05 | 0.07 ± 0.06 | 0.25 ± 0.17 | 0.5 | ≈ 0 |
| 11. | Minor axis length AI | 0.05 ± 0.04 | 0.06 ± 0.05 | 0.2 ± 0.15 | 0.18 | ≈ 0 |
| 12. | Aspect Ratio AI | 0.08 ± 0.06 | 0.08 ± 0.06 | 0.13 ± 0.1 | 0.67 | ≈ 0 |
| 13. | Eccentricity AI | 0.04 ± 0.04 | 0.04 ± 0.04 | 0.23 ± 0.27 | 0.71 | ≈ 0 |
| 14. | Equivalent diameter AI | 0.03 ± 0.03 | 0.04 ± 0.05 | 0.22 ± 0.15 | 0.75 | ≈ 0 |
| 15. | Orientation AI | 0.12 ± 0.12 | 0.12 ± 0.09 | 0.01 ± 0.85 | 0.89 | ≈ 0 |
| 16. | Roundness AI | 0.05 ± 0.04 | 0.06 ± 0.05 | 0.09 ± 0.08 | 0.97 | ≈ 0 |
| 17 | $p_{00}$ | 0.93 ± 0.02 | 0.93 ± 0.02 | 0.92 ± 0.03 | 0.13 | 0.15 |
| 18 | $p_{10}$ | -0.01 ± 0.02 | -0.01 ± 0.02 | 0.01 ± 0.02 | 0.55 | ≈ 0 |
| 19 | $p_{01}$ | -0.09 ± 0.02 | -0.09 ± 0.02 | -0.03 ± 0.02 | 0.92 | ≈ 0 |
| 20 | $p_{20}$ | -0.11 ± 0.01 | -0.11 ± 0.01 | -0.07 ± 0.01 | 0.48 | ≈ 0 |
| 21 | $p_{11}$ | -0.11 ± 0.02 | -0.11 ± 0.02 | -0.03 ± 0.02 | 0.65 | ≈ 0 |
| 22 | $p_{02}$ | -0.11 ± 0.01 | -0.11 ± 0.01 | -0.07 ± 0.02 | 0.65 | ≈ 0 |
| 23 | $p_{30}$ | 0.01 ± 0.01 | 0.01 ± 0.01 | 0 ± 0.01 | 0.2 | ≈ 0 |
| 24 | $p_{21}$ | 0.03 ± 0.01 | 0.02 ± 0.01 | 0.01 ± 0.01 | 0.49 | ≈ 0 |
| 25 | $p_{12}$ | 0 ± 0.01 | 0 ± 0.01 | -0.01 ± 0.01 | 0.97 | ≈ 0 |
| 26 | $p_{03}$ | 0.03 ± 0.01 | 0.03 ± 0.01 | 0.01 ± 0.01 | 0.89 | ≈ 0 |
| 27 | SE | 0.05 ± 0.03 | 0.06 ± 0.04 | 0.01 ± 0.01 | 0.19 | ≈ 0 |
| 28 | $R^2$ | 0.96 ± 0.02 | 0.96 ± 0.02 | 0.98 ± 0.01 | 0.26 | ≈ 0 |
| 29. | $R^2_{adj}$ | 0.96 ± 0.02 | 0.95 ± 0.03 | 0.98 ± 0.01 | 0.26 | ≈ 0 |
| 30. | RMSE | 0.02 ± 0.01 | 0.02 ± 0.01 | 0.01 ± 0 | 0.26 | ≈ 0 |
| 31. | Caudate SBR | 2.97 ± 0.62 | 2.85 ± 0.57 | 2.02 ± 0.54 | 0.26 | ≈ 0 |
| 32. | Putamen SBR | 2.13 ± 0.56 | 2.06 ± 0.48 | 0.84 ± 0.31 | 0.55 | ≈ 0 |
| 33. | Caudate SBR AI | 0.08 ± 0.06 | 0.08 ± 0.06 | 0.18 ± 0.12 | 0.66 | ≈ 0 |
| 34. | Putamen SBR AI | 0.11 ± 0.09 | 0.12 ± 0.1 | 0.37 ± 0.25 | 0.67 | ≈ 0 |

$p1$ and $p2$ represent the $p$-values of features for healthy normal vs. SWEDD comparison and for early PD vs. (healthy normal/SWEDD) comparison, respectively. All the features, except the model coefficient $p_{00}$, are highly statistically significant ($p$-value<0.01) in depicting the changes in PD as compared to normal or SWEDD. On the other hand, no feature is significant ($p$-value>0.05) while comparing normal and SWEDD groups.

We observe the following from the table:

a) Area, major axis length, minor axis length and equivalent diameter decreases in PD as compared to the non-degenerative groups (healthy normal or SWEDD). This indicates that the size of uptake regions reduces in PD.

b) Aspect ratio and eccentricity decreases in PD, becoming close to 1 and 0, respectively. Roundness increases, becoming close to 1 in PD. This indicates that the uptake region becomes more 'circular' or 'dot' shaped in PD.

c) Orientation decreases in degenerative PS. Its value of 49.46 ± 7.89 degrees for healthy normal or 50.75 ± 7 degrees for SWEDD indicate normal uptake in both posterior and anterior striatum. However, during PD, its value decreases to 29.66 ± 17.34 degrees due to the loss of activity that proceeds from posterior to anterior striatum. This is consistent with the clinical perspective of PD.

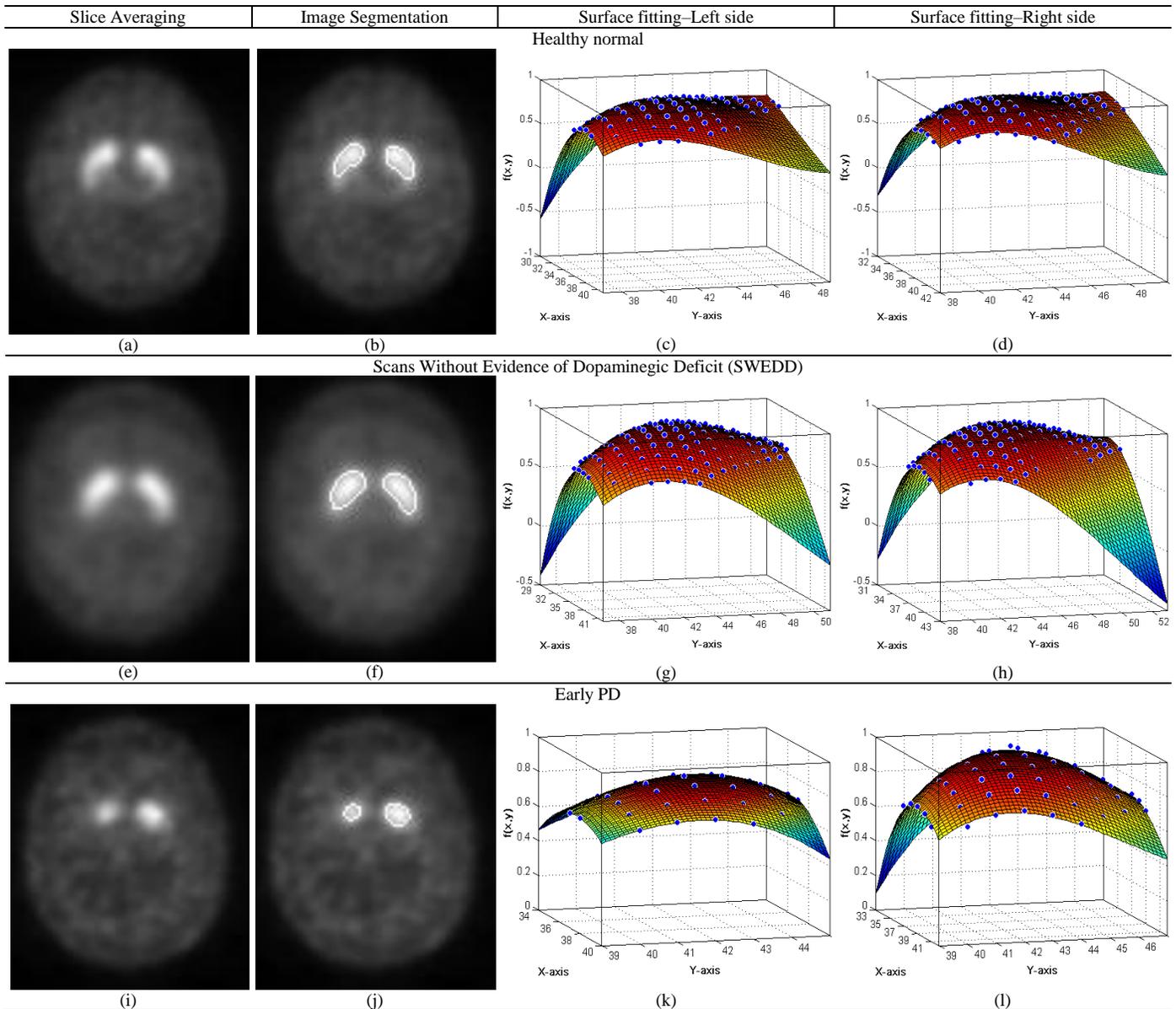

| Slice Averaging | Image Segmentation | Surface fitting–Left side | Surface fitting–Right side |
|---|---|---|---|

Healthy normal

(a)     (b)     (c)     (d)

Scans Without Evidence of Dopaminegic Deficit (SWEDD)

(e)     (f)     (g)     (h)

Early PD

(i)     (j)     (k)     (l)

Fig.2. Slice averaging, image segmentation and surface fitting for healthy normal (a, b, c, d), SWEDD (e, f, g, h), and early PD (i, j, k, l). The shape of the uptake region undergoes a change from a 'comma' shaped to more 'circular' or 'dot' shaped during PD as compared to normal or SWEDD. The fitted surfaces also show variation during PD as compared to the other groups. No substantial difference is observed between normal and SWEDD groups visually. The figures 2(c, d, g, h, k, l) show the fitted surfaces with positive (peaks) and negative (valleys) curvatures based on the intensity values. It is a 3D representation of the intensity values with $x$ and $y$ axis representing the image coordinates. The rectangle (or the cuboid in 3D) represents the bounding box of the segmented region. Cubic surfaces are fitted separately for the two segmented regions (left and right striatum). The left side is kept as the reference for the fitting process. Therefore, before carrying out the fitting process for the right side, the image is flipped right to left. This is carried out so that the coordinates for the left and right are in same scale. The top left corner of the image is (0,0) and the bottom right is (91,109). The pixel size is 2 mm × 2 mm.

d) Coefficients corresponding to linear terms ($p_{10}$ and $p_{01}$) and quadratic terms ($p_{11}$, $p_{20}$ and $p_{02}$) in the cubic model are higher, and coefficients corresponding to cubic terms ($p_{30}$, $p_{21}$, $p_{12}$ and $p_{03}$) are lower in PD as compared to the other groups. This indicates that the surface corresponding to PD is close to quadratic in nature. The constant term ($p_{00}$) showed little or no difference between the three groups. This is because $p_{00}$ essentially reflects the height of the curve from the ground (or base surface), and the height which is dependent on the overall intensity distribution is similar between the groups as it is normalized. $p_{00}$ has no effect on the curvature of the surfaces. The goodness-of-fit measures showing higher $R^2$ or $R^2_{adj}$, and lower $SE$ or $RMSE$ implies that there is higher degree of fitting in early PD as compared to other groups.

Table I also shows the results of statistical testing of these features, for healthy normal vs. SWEDD comparison and PD vs. Healthy Normal/SWEDD group comparison. None of the features showed statistical significance ($p$-value>0.05) while comparing healthy normal with SWEDD.

The basic aim of the study is in discriminating degenerative PS from the non-degenerative scans through shape and surface fitting in SPECT imaging. As there is no substantial difference between the healthy normal and SWEDD groups, as observed from the box plots and statistical analysis (Table I), we consider the two groups to form a single Normal/SWEDD group for classification modeling process. While comparing the early PD group and this combined Normal/SWEDD group, we observed that all features, except the model coefficient $p_{00}$, showed high statistical significance ($p$-value<0.05) indicating their usefulness in discriminating degenerative PS.

### B. Classification modeling

Of all the features, 29 features (all features except the model coefficient $p_{00}$) that are statistically significant are used for subsequent classification modeling. Table II shows the performance measures obtained for the various classifiers used. It is observed that all classifiers performed with a high accuracy. SVM classifier (using Radial Basis Function (RBF) kernel) gave the highest accuracy (slightly higher than other classifiers) and area under the ROC curve (AUC) of 97.29% and 99.26%, respectively. The parameters $C$ and $\gamma$ for SVM were obtained using 10-fold cross validation (CV) as 1 and 0.0625, respectively.

In our previous work [24], we had showed that SVM classifier using SBR features produced an accuracy of 96.14% that was higher than the state-of-the-art studies. In this work, we observe that SVM classifier (or any other classifier used in the study) using shape- and surface fitting-based features gives higher accuracy than our previous work. A more detailed comparison is given in Sec III.E.

In the boosted trees model, the minimum size of parent node and leaf node is specified as 10 and 5, respectively. The number of trees in the model is chosen as 70 based on the observation that at this point the 10-fold CV error was smallest and the error was almost same after this point. The number of trees in the Random forests model is chosen as 65 based on lower 10-fold CV and out-of-bag errors.

### TABLE II
PERFORMANCE MEASURES OBTAINED FOR THE CLASSIFIER USED.

| Performance measures | SVM | Boosted Trees | Random Forests | Naïve Bayes |
|---|---|---|---|---|
| Accuracy | $97.29 \pm 0.11$ | $96.76 \pm 0.23$ | $96.90 \pm 0.17$ | $96.88 \pm 0.09$ |
| Sensitivity | $97.37 \pm 0.10$ | $97.09 \pm 0.25$ | $97.18 \pm 0.23$ | $96.43 \pm 0.14$ |
| Specificity | $97.18 \pm 0.22$ | $96.29 \pm 0.42$ | $96.49 \pm 0.32$ | $96.47 \pm 0.16$ |
| AUC | $99.26 \pm 0.06$ | $99.16 \pm 0.12$ | $99.08 \pm 0.11$ | $98.99 \pm 0.07$ |

### C. Estimation of importance of features

Feature importance is carried out to observe the relative importance of the features, and to compare the shape-based and surface fitting-based features as computed in our study, with the standard SBR-based features (computed from SBR values that are calculated by experts at PPMI and obtained from the PPMI database). Fig. 3 shows the plot of feature importance scores for each feature.

The number of trees in the Random forest model is chosen as 75 based on lower 10-fold CV error and out-of-bag error. Major axis length, model coefficient $p_{11}$ and mean putamen SBR are observed to be features of higher importance. Our observation of mean putamen SBR being an important feature is consistent with previous studies [30, 31] which show that, during PD, greater reduction occurs in the putamen than in the caudate. Major axis length, which is a feature that reflects the spread of the dopaminergic activity, decreases in PD. This is consistent with the observation by Staff et al.[15] that, as the deterioration progresses, deficit in activity proceeds from posterior to anterior striatum which results in a decrease of the region of activity. Model coefficient $p_{11}$ reflects the curvature of the surfaces. Higher the absolute value of this coefficient, more negative the surface curvature. Its value is close to 0 indicating less negative curvature or more positive curvature in PD. The loss of curvature in PD is due to the loss of activity in the posterior striatum (or the putamen) which is consistent with the Staff et al.[15] study.

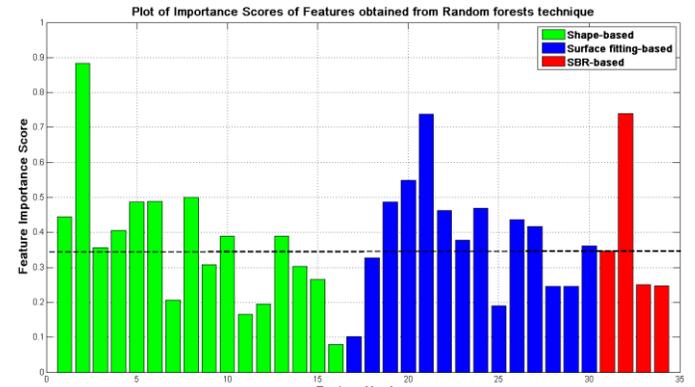

Fig. 3. Plot of feature importance scores for all features: shape-based (features from 1 to 16), surface fitting-based (features from 17 to 30) and SBR-based features (features from 31 to 34). $x$-axis represents features indicated by their corresponding numbers as given in Table I. Major axis length (feature number 2), model coefficient $p_{11}$ (feature number 21) and mean putamen SBR (feature number 32) are observed to be the most important features. This observation is consistent with the clinical perspective of the deterioration process in degenerative PS such as in PD. The bold dashed line in the figure corresponds to the score of the mean caudate SBR which has the second highest feature importance score among the SBR-based features. 10 shape-based and 10 surface fitting-based features show higher scores than the mean caudate SBR. Random forests, which was used for studying variable importance, also computes the predictive power using the out-of-bag observations. They are as follows: Accuracy=97.07%, Sensitivity = 97.32% and Specificity = 96.69%.

Mean caudate SBR (feature 31 in Fig. 3) has the second highest score among the SBR-based features. To compare shape- and surface fitting-based features with SBR-based features, we see the relative importance of both with the mean caudate SBR. Nine among the 16 shape-based features and nine among the 14 surface fitting-based features have higher scores than the mean caudate SBR. It indicates that the shape- and surface fitting-based features show higher discriminatory power and has the potential in distinguishing scans with deficit from scans without deficit.

Along with estimating the feature importance, Random forests technique also provides an average out-of-bag error which is an unbiased estimator of the true ensemble error and an estimate of the predictive power. We observe that using SBR-based features along with shape-based and surface fitting-based features did not substantially improve the classification performances (Accuracy difference = 0.17%). This indicates that the shape- and surface fitting-based features contained enough information essential for classification, and hence, these features have the potential to be useful in a clinical setting for the diagnostics of PD.

In the limitations, the present approach does not involve another main clinical scenario, the differentiation of other neurodegenerative parkinsonisms (differential diagnosis), or address issues of vascular change affecting dopamine transporters.

### D. Note on slice selection and threshold selection

Our slice selection is based on the Society of Nuclear Medicine (SNM) recommendations which mention that at least 3 consecutive slices in the target region are to be used—those with the highest activity [16], and within the same center, the number of slices chosen should be kept consistent. For illustration, we computed the uptake areas for slices from 35 to 48 (rest of the slices did not show significant striatal activity) as shown in Fig. 4. For our study, we selected slices numbered from 35 to 48 and taken their average for further analysis. This number was selected based on careful empirical experiments making it very less machine dependent.

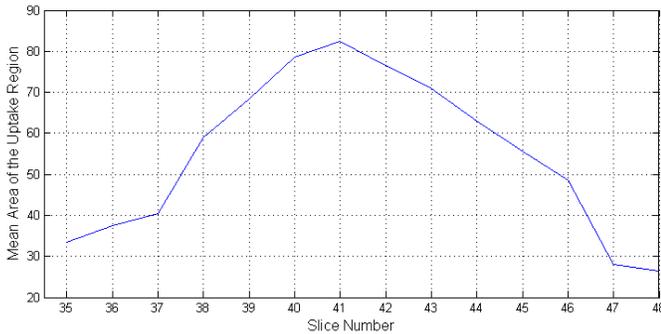

Fig. 4. Plot of the areas (in pixels) of the segmented striatal regions from slices 33 to 50 (from the total 91 slices).

Our method is not fully automatic, like the classifiers, with regard to the threshold selection. Although a threshold is applied to the images for segmentation, it is not totally subjective as well. This is due to the normalization process that is carried out before segmentation. The segmented regions are carefully assessed by expert investigator. The means ±SD values of the thresholds used for healthy normal, SWEDD and early PD are 0.63±0.04, 0.63±0.03 and 0.69±0.05, respectively. The plot of histograms of thresholds used for the three groups are shown in Figs. 5 (a, b & c), respectively. It is important to note that the variability of the thresholds used for each group is very low as observed from the very low standard deviations, 0.04 (6.34 %), 0.03 (4.76%) and 0.05 (7.24%) for healthy, SWEDD and early PD groups, respectively.

The thresholds used for the PD case is higher due to the following. During PD, dopamine transporters (DATs) are first lost in the putamen (lower portion of the segmented regions or striatum) which correspond to the lower range of intensities in the high activity striatal region, then in the caudate (upper portion of the segmented regions or the striatum), or in other words, it proceeds from the posterior to anterior striatum. Due to this, there is loss of the negative curvature regions which correspond to the posterior striatum.

A similar study by Staff et al.[15], where they carry out segmentation of the striatal regions using a threshold and then quantify its shape using a shape feature. They analyzed the reproducibility in terms of inter- and intra-observer variability, and observed a good inter- and intra-observer reproducibility.

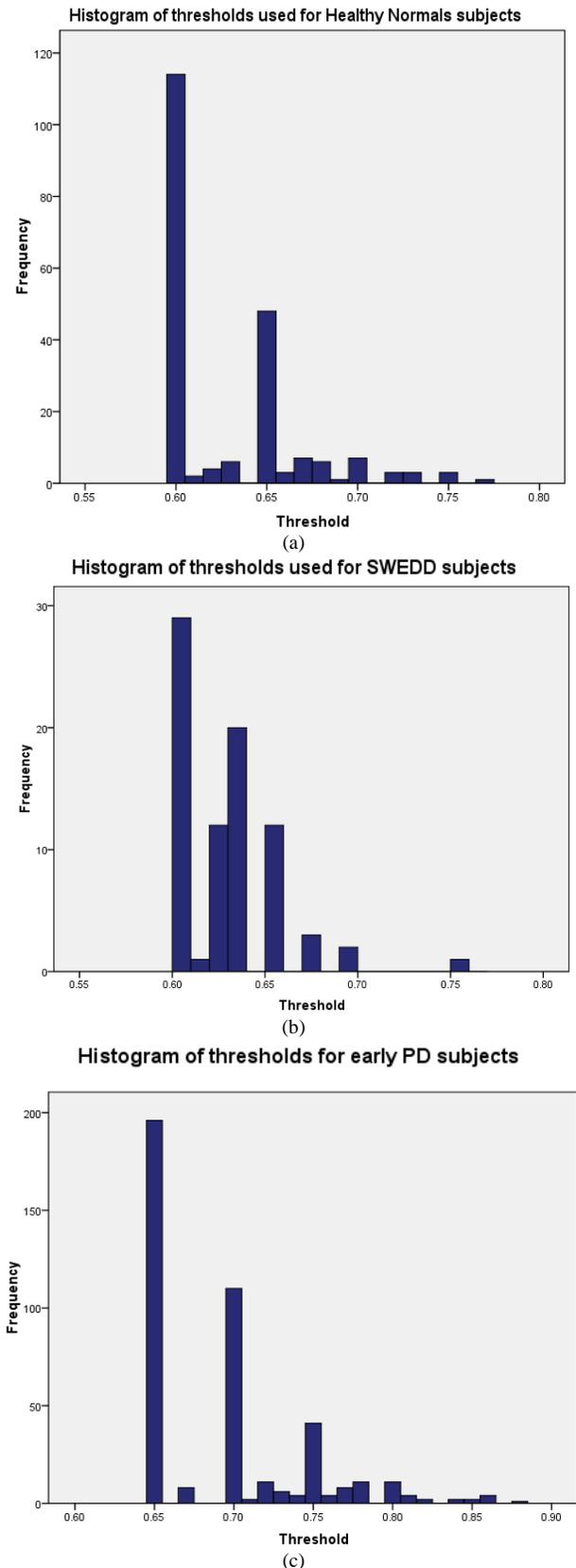

Fig. 5. Histogram of the thresholds used for segmenting the striatal region in (a) healthy normal, (b) SWEDD and (c) Early PD subjects. The thresholds used for healthy normal, SWEDD and early PD were 0.63±0.04, 0.63±0.03 and 0.69±0.05, respectively.

### E. Comparison with related works

We discuss the difference in performance and approaches between the method used in this study and related works with the help of a table as shown below (Table III).

| Study | Sample size | Method | Accuracy/ AUC |
|---|---|---|---|
| Segovia et al. [11] | 95 PS, 94 N | Extracted voxels from the striatum and performed data decomposition using partial least squares, followed by classification using SVM. | 94.7% |
| Illan et al. [12] | 100 PS, 108 N | Used voxels of the complete brain as features and then performed classification using a SVM with linear kernel. | 96.81% |
| Rojas et al. [13] | 41 PS, 39 N | Obtained best performance with the Principal Component Analysis derived features from the high intensity voxels of the striatum and classification using SVM | 95% |
| Towey et al. [14] | 79 PS, 37 control | Used Singular Value Decomposition (SVD) to extract significant voxels, followed by classification using Naïve Bayes | 94.8% |
| Prashanth et al. [24] | 369 PD, 179 N | Used striatal binding ratio values as features, followed by classification using SVM | 96.14% |
| Staff et al. [15] | 27 PS, 25 control | Segmented high uptake areas of striatum, and quantified its shape via the aspect ratio of the ellipse that was fitted to the region | 94% |
| Oliveira et al. [32] | 445 PD, 209 N | Used voxels from the striatum as features and then performed classification using a SVM classifier | 97.86% |
| Martinez-Murcia et al. [33]* | 158 PD, 111 N | Computed Haralick texture features via a gray level co-occurrence matrix from the brain voxels and used SVM classifier with linear kernel. | 97.4% |

\* study used three different databases and obtained different accuracies. The table shows the highest accuracy.

Our results are highly competitive when compared to related works. It is to be noted that different databases have been used in different studies which may bias the comparison. The main take away from the present study is that the analysis gave high performance using a large database, the PPMI, which is one of the large-scale and standard databases publicly available for early PD. It is encouraging to observe high performance from other studies as well which implies the potential of quantification followed by machine learning in SPECT imaging for the diagnosis of PD. However, we would like to point out that the related works used leave-one-out cross validation (LOOCV), which is well known to suffer from high variances, for estimating the performance of the classifiers. In LOOCV where one sample is used for testing and the rest for training, tends to select models with higher variances, which may lead to overfitting. In our approach, we carry out repeated 10-fold cross validation, as recommended by [34], which has lower variance, and therefore tend to give more stable models. Also, most of the related works had a limitation of smaller database in their study.

### F. Future of DaTSCAN on SWEDD

Distinguishing SWEDD from PD is important as most of the SWEDD subjects receive unnecessary and inappropriate treatment, with huge side-effects, for many years. DaTSCAN has shown huge potential in detecting SWEDD. In a number of clinical trial studies in early PD, using SPECT imaging as the secondary outcome measure, has observed that about 10-15% of subjects with the clinical diagnosis of PD had dopaminergic scans without evidence of dopaminergic deficit [2, 8, 35]. Substantial evidence in terms of long-term follow-up of these subjects indicated poor response to levodopa and lack of progression on sequential dopaminergic imaging [6, 9]. These suggest that most of these patients do not have involvement in the nigrostriatal pathway and do not have PD, indicating that this is an issue of misdiagnosis rather than inadequate sensitivity of the scan.

A study done by Schwingenschuh et al. [7] observed that adult-onset dystonia is a possible underlying diagnosis for SWEDD, rather than PD. Catafau et al. [4] performed a study to investigate the clinical impact of [123]I-Ioflupane SPECT in patients with clinically uncertain PS. And they observed that after imaging, diagnosis was changed in 52% (61 out of a total 118) of patients. All patients with a final diagnosis of presynaptic PS had an abnormal image, whereas 94% of patients with nonpresynaptic PS had a normal scan. Imaging increased confidence in diagnosis, leading to changes in clinical management in 72% of patients. They also examined the relationship between final diagnosis and imaging result, [123]I-Ioflupane SPECT imaging had an important impact on the final diagnosis, by the finding that 100% of patients with a final diagnosis of presynaptic PS had an abnormal image result, whereas 94% of patients with a final diagnosis of nonpresynaptic PS had a normal image result. [123]I-Ioflupane SPECT is therefore a recommended adjunct to the diagnosis of patients with uncertain parkinsonism (where there is diagnostic uncertainty), especially SWEDD.

## IV. CONCLUSION

Accurate differential diagnosis of PD from the non-degenerative PS, tremor disorders or SWEDD cases in their early stages is a challenging and important problem. As these conditions share many common symptoms, it is a source for misdiagnosis. Accurate identification of degenerative PS from other non-degenerative variants is crucial for effective patient management. In our work, we process SPECT images of healthy normal, early PD and SWEDD, and carry out shape analysis and surface fitting to compute discriminatory features. We observe that the computed shape-based and surface fitting-based features show significant variation between scans showing dopaminergic deficit from scans which did not. The classification models developed using these features performed with a high accuracy, sensitivity and specificity. It is inferred from the study that shape analysis and surface fitting are useful approaches to develop classification models that can aid a clinician in quantitatively observing the deterioration and thereby, aiding in the diagnostic process.

### ACKNOWLEDGEMENT

PPMI, a public-private partnership, is funded by the Michael J. Fox Foundation for Parkinson's Research and other funding partners include AbbVie, Avid Radiopharmaceuticals, Biogen Idec, Bristol-Myers Squibb, Covance, GE Healthcare,

Genentech, GlaxoSmithKline, Eli Lilly and Company, Lundbeck, Merck & Co., Meso Scale Discovery, Pfizer, Piramal, Hoffmann-La Roche, and UCB (Union ChimiqueBelge). The authors also thank Dr. Michael Fox, MD, PhD, at the Beth Israel Deaconnes Medical Center, a teaching affiliate of the Harvard Medical School for his valuable suggestions. The authors thank Dr. Jeevanand, post doctoral fellow, Nanyang Technological UniversitV.VI.VII.y for helping with the revision.